\documentclass[%
 reprint,
 amsmath,amssymb,
 aps,
]{revtex4-2}

\usepackage{graphicx}
\usepackage{dcolumn}
\usepackage{bm}

\begin{document}

\preprint{APS/123-QED}

\title{Exploring the Effect of the Number of Hydrogen Atoms on the
Properties of Lanthanide Hydrides by DMFT}

\author{Yao Wei}
\affiliation{King's College London, Theory and Simulation of Condensed Matter
(TSCM), The Strand, London WC2R 2LS, UK}
\author{Elena Chachkarova}
\affiliation{King's College London, Theory and Simulation of Condensed Matter
(TSCM), The Strand, London WC2R 2LS, UK}
\author{Evgeny Plekhanov}
\affiliation{King's College London, Theory and Simulation of Condensed Matter
(TSCM), The Strand, London WC2R 2LS, UK}
\author{Nicola Bonini}
\affiliation{King's College London, Theory and Simulation of Condensed Matter
(TSCM), The Strand, London WC2R 2LS, UK}
\author{Cedric Weber}
\affiliation{King's College London, Theory and Simulation of Condensed Matter
(TSCM), The Strand, London WC2R 2LS, UK}

\date{\today}

\begin{abstract}
Lanthanide hydrogen-rich materials have long been considered as one of the candidates with high-temperature superconducting properties in condensed matter physics, and have been a popular topic of research. Attempts to investigate the effects of different compositions of lanthanide hydrogen-rich materials are ongoing, with predictions and experimental studies in recent years showing that substances such as LaH$_{10}$, CeH$_{9}$, and LaH$_{16}$ exhibit extremely high superconducting temperatures between 150--250 GPa. In particular, researchers have noted that, in those materials, a rise in the $f$ orbit character at the Fermi level combined with the presence of hydrogen vibration modes at the same low energy scale will lead to an increase in the superconducting transition temperature. Here, we further elaborate on the effect of the ratios of lanthanide to hydrogen in these substances with the aim of bringing more clarity to the study of superhydrides in these extreme cases by comparing a variety of lanthanide hydrogen-rich materials with different ratios using the dynamical mean-field theory (DMFT) method, and provide ideas for later structural predictions and material property studies.
\begin{description}
\item[Keywords]
superconductivity; elemental composition; high pressure; DMFT
\end{description}
\end{abstract}

\maketitle


\section{\label{sec:level1}Introduction}

Recently, hydrogen-rich materials have attracted great attention as potential candidates for high-temperature superconductivity, which is referred to as the holy grail of condensed matter physics. Since hydride causes high-frequency vibrations of hydrogen atoms at high pressure, it leads to stronger electron-phonon coupling, making it a potential high-temperature superconducting material \cite{mcmahon2011high,borinaga2016anharmonic}.
Almost 90 years ago, theoretical physicists E. Wigner and H. B. Huntington predicted that hydrogen, the lightest element in nature, would be extremely likely to metalize in a high pressure environment \cite{wigner1935possibility}.
An early study in the 1960s on metallic hydrogen found superconductivity at temperatures close to or even above room temperature, which suggests that  $T_c$ of molecular hydrogen under high pressure is about 100--240 K, and that $T_c$ of atomic hydrogen can reach 300--350 K also under high pressure \cite{cudazzo2008ab,mcmahon2012properties}. However, it is currently still difficult to perform relevant experiments due to the stringent synthesis requirements and ultra-sensitivity of metallic hydrogen under high pressure \cite{eremets2019semimetallic,loubeyre2020synchrotron}. As a result of development of computational tools for predicting crystal structures and calculating properties of materials, ultra-high pressure generators have emerged. Nowadays, theoretical and experimental research on hydrogen-rich compounds covers binary hydride materials formed by almost all elements, which also includes the lanthanum-based high-temperature superconducting material LaH$_{16}$~\cite{sun2020second}, as well as CeH$_{9}$ \cite{weber2022computational, wang2021effect,li2019predicted}, CeH$_{16}$, and YbH$_{10}$ \cite{wei2022high,kruglov2020superconductivity}. Research on these materials has found that binary lanthanide superconducting materials with high hydrogen content at high pressure are characterized by high symmetry and that they exhibit a space group structure of hexagonal and cubic crystal systems under 200--300 GPa. In addition to lanthanide superconducting materials, there are also hydrogen-rich superconducting materials with yttrium elements, which are chemically similar to lanthanides and often classified as rare earth metals YH$_{6}$, YH$_{9}$, and YH$_{10}$ \cite{kong2021superconductivity, struzhkin2020superconductivity}. Among the transition elements, many actinide-related superconducting materials have also been found, such as AcH$_{10}$, AcH$_{12}$, AcH$_{16}$, and ThH$_{9-10}$ \cite{semenok2018actinium, semenok2020superconductivity, kvashnin2018high}.

Furthermore, some common metal elements have been found to form hydrogen-rich materials with superconducting properties, which include CaH$_{6}$ \cite{wang2012superconductive,shao2019unique,ma2021high,li2021superconductivity},
MgH$_{6}$ \cite{szczesniak2016superconductivity}, MgH$_{12}$~\cite{lonie2013metallization}, to name a few.

In addition to the hydrogen-rich superconducting materials containing metallic elements, hydrogen-rich materials with nonmetals have also attracted much attention. In recent research, Ranga P. Dias reported the superconductivity of a C-S-H system with a maximal superconducting transition temperature of 287.7 ± 1.2 K (approximately 15 degrees Celsius) at 267 ± 10 GPa \cite{snider2020room}. Besides, it was predicted that $T_{c}$ values of PbH$_{8}$, SbH$_{4}$, and AsH$_{8}$ would be 178 K, 150 K, and 100 K above 200 Gpa, 350 GPa, and 150 GPa respectively, as covalent hydrogenates~\cite{chen2021phase,fu2016high}.

It has been found that LaH$_{10}$ with face-centered cubic structure is a good metal, forming high electron-density energy bands at the Fermi level. The finite occupancy of the La-$4f$ orbitals, which are successively pushed at the Fermi level as the external pressure increases~\cite{fedorov1993surface}, ensures a particularly high $T_{c}$ in La-H systems. The localized La-4$f$ orbitals become more stable compared to the La-6$s$ and La-5$d$ orbitals due to the pressure of the external environment, which leads to new prominent quantum states originating from strong electronic correlations \cite{fedorov1993surface}.

Therefore, in order to more accurately describe the properties and physical laws of hydrogen-rich high-temperature superconductors at high pressures, the quantitative theoretical calculations were used in our research. The interaction between the localized electrons and the strong lattice vibrations of the hydrogen atoms facilitates superconductivity in these materials. A thorough description of the electrical properties is required for a precise analysis of this interaction, which brings notorious difficulties for $f$-systems correlation, as it requires that both mobile and localized electrons are treated on the same basis \cite{pfleiderer2009superconducting}.

Here, we present a study of the superconducting properties of two members of lanthanum hydrates family LaH$_{10}$ and LaH$_{18}$ in order to learn the effect of the hydrogen content on T$_c$ and to find out if the many-body corrections to the DFT results are equally important for lanthanides with different hydrogen content.
We employ a practical first-principles computationally consistent platform that uses a many-body correction of electron spectral weights and then feeds back into $T_{c}$'s estimate. In the present work, we limit ourselves to the DMFT  corrections of the electronic spectrum only.

This paper proves that many-body correction in $f$-orbital systems may lead to changes occurring in spectral weights on the order of an eV. Our research presents an assessment of the physical properties of two La-H systems for different quantities of hydrogen atoms, paying particular attention to the influence of related factors on the spectral properties. It was found that LaH$_{10}$ and LaH$_{18}$ have stable crystal structures at pressures up to 400 GPa, while LaH$_{18}$ is unstable at lower pressure; hence, we focus this investigation on modeling the materials at 400 GPa, as our aim is to compare the correction validity for different hydrogen compositions.

\section{Discussion}

Electron-phonon interactions are thought to be the origin of superconductivity in lanthanide hydrides. According to the Migdal--Eliashberg theory, there are four factors to define $T_{c}$: the characteristic phonon frequency $\omega_{\log}$, the electron-phonon coupling strength $\lambda$, the density of states at the Fermi level $N(\epsilon_F)$, and the Coulomb pseudopotential, $\mu^{\star}$. Density functional theory (DFT), which uses standard pseudo-potentials and exchange correlation functionals, e.g., PBE, to explain lattice dynamics, is universally considered to be correct. It is generally acknowledged that DFT struggles to describe systems possessing strong electronic correlations with strongly localized $d$ and $f$ shells, and the use of many-body corrections are necessary.

\begin{figure*}
\includegraphics[width=1\textwidth]{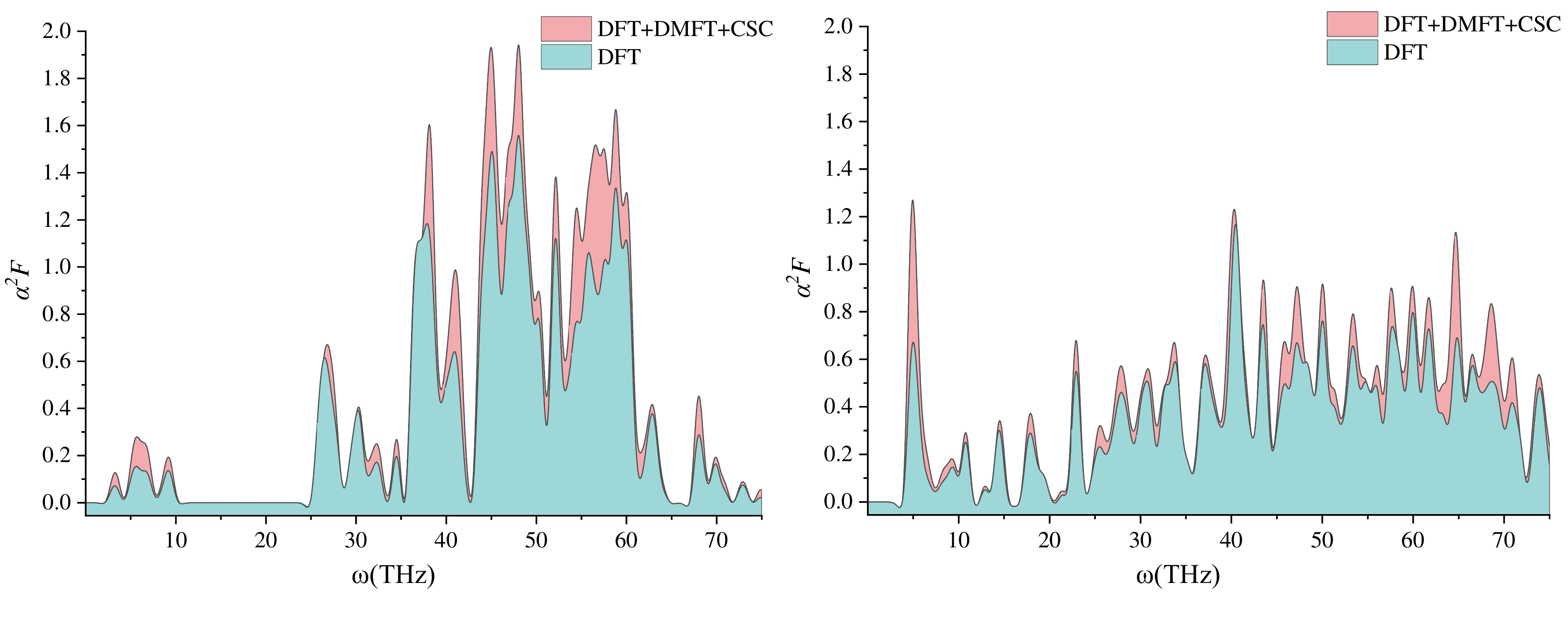}
\caption{Eliashberg function $\alpha^2F(\omega)$ for (a) LaH$_{10}$ and (b)
LaH$_{18}$; the spectral weight at the Fermi level is calculated with different extents of approximation: (i) DFT and (ii) with the full-charge self-consistent formalism (DFT+DMFT+CSC).}
\label{a2f}
\end{figure*}

\begin{figure*}
\includegraphics[width=\textwidth]{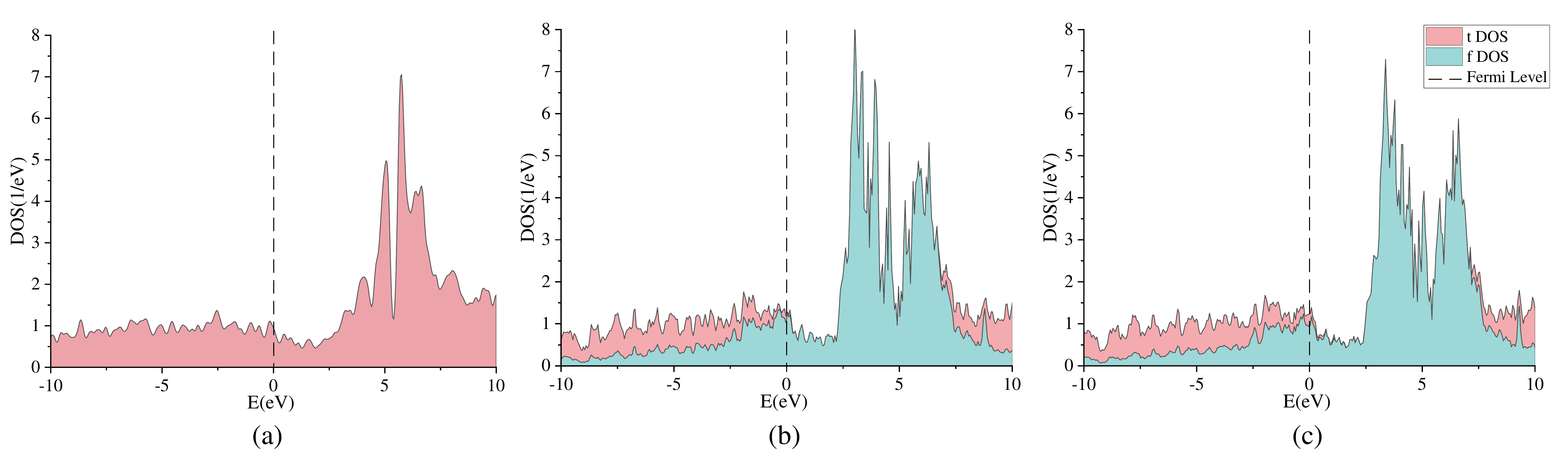}
\caption{(\textbf{a}) Density of states obtained by DFT calculations. $t DOS$ and $f DOS$ denote the density of states corresponding to the lattice and $f$ impurity Green's function, respectively. In (\textbf{b},\textbf{c}), we show the density of states, obtained by the one-shot DFT+DMFT and the full-charge self-consistent DFT+DMFT+CSC, respectively. All calculations were performed in the Fm3m phase of LaH$_{10}$ at 400 GPa.}
\label{LaH10dos}
\end{figure*}

\begin{figure*}
\includegraphics[width=\textwidth]{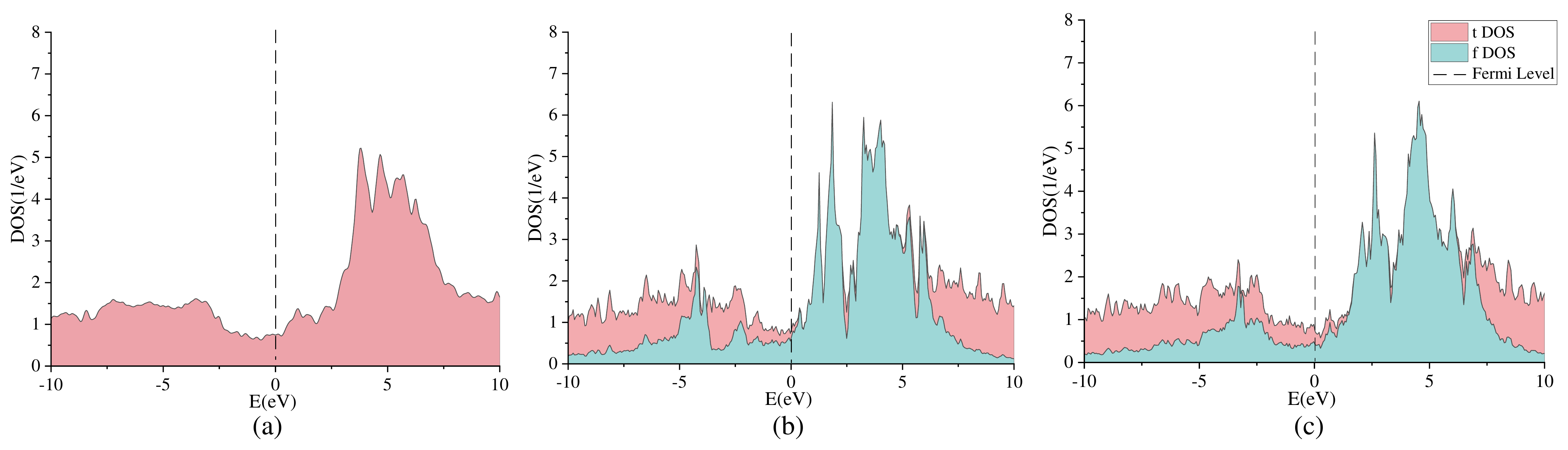}
\caption{(\textbf{a}) Density of states obtained by DFT calculations. $t DOS$ and $f DOS$ denote the density of states corresponding to the lattice and $f$ impurity Green's function, respectively. In (\textbf{b},\textbf{c}), we show the density of states, obtained by the one-shot DFT+DMFT and the full-charge self-consistent DFT+DMFT+CSC, respectively. All calculations were performed in the Fm3m phase of LaH$_{18}$ at 400 GPa.}
\label{LaH18dos}
\end{figure*}

In the following analysis, we used DFT combined with DMFT. In order to capture the local paramagnetic moment in lanthanide elements, one has to correct for local charge and spin fluctuations captured by DMFT. Changes in the orbital character at the Fermi surface produced by spectral weight transfer linked to Hubbard $f$ band splitting are accounted for using DMFT. In accordance to the Allen--Dynes formalism, this affects low-energy electron-electron scattering processes via transfer of phonon momentum, which will be indicated in the following section.

\begin{figure}
\includegraphics[width=1\columnwidth]{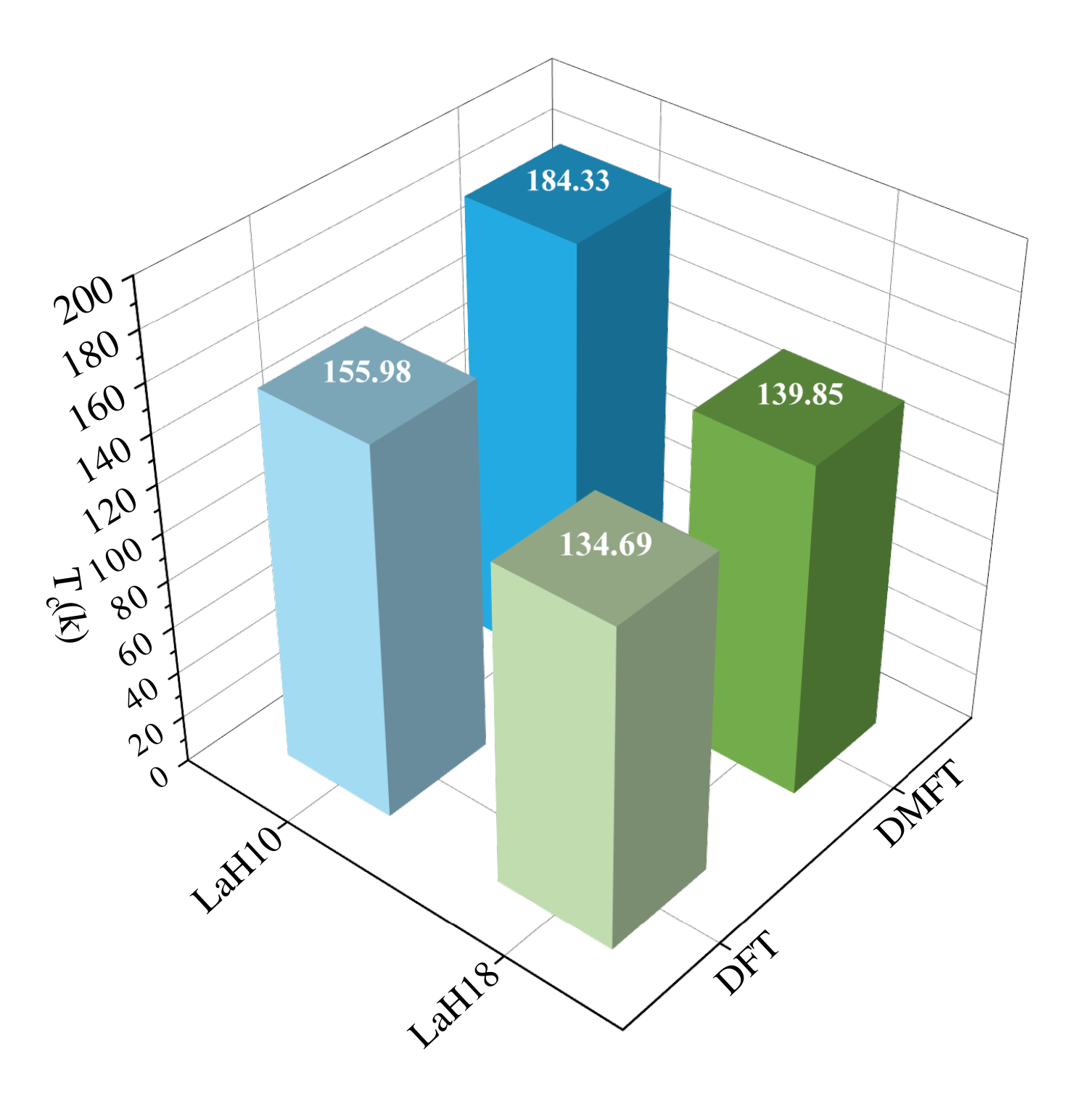}
\caption{The superconducting temperature $T_{c}$ obtained by the Allen and Dynes formalism for LaH$_{10}$ and LaH$_{18}$ at 400 GPa. We obtained a theoretical estimate for LaH$_{10}$ of $T_c=155.98$ K by DFT and $T_c=184.33$ K by DMFT, as well as a theoretical estimate for LaH$_{18}$ of $T_c=134.69$ K by DFT and \mbox{$T_c=139.85$ K} by DMFT. We have used values of $U=6$ eV and $J=0.6$ eV. All calculations were performed in the 
Fm3m-LaH$_{10}$ and Fmmm-LaH$_{18}$ at 400 GPa.}
\label{Tc}
\end{figure}

We compared the effect of different DMFT electronic charge self-consistency schemes. A comparison is shown in Figure~\ref{a2f} between (i) PBE DFT and (ii) DFT+DMFT with full charge self-consistent formalism (DFT+DMFT+CSC). According to our research, we found that the Eliashberg function, $\alpha^2F(\omega)$, was significantly influenced by the correlation effects, which is defined in Equation (2). We utilized the Koster--Slater interaction vertex for the La correlated manifold, with values of $U=6$ eV and  $J=0.6$ eV. From Figure~\ref{Tc} we can conclude that the DFT+DMFT+CSC method enables an increase in the superconducting temperature, which confirms the large contribution of the many-body effect to the prediction of the superconducting temperature of lanthanide hydrides, and justifies the need for more detailed correction models.

We noted that for the higher composition of hydrogen, LaH$_{18}$, the correction to the critical temperature is less significant compared to the lower hydrogen composition case. Figure ~\ref{a2f} shows that the latter has a few well determined peaks that are enhanced by the DMFT+CSC correction, whilst in the LaH$_{18}$ case, the Eliashberg function has a more evenly spread character with lower height of the peaks. Consequently, for the many-body corrections, we obtained an approximately $18\%$ correction to LaH$_{10}$ critical temperature compared to just under $4\%$ for LaH$_{18}$, as can be seen from Figure ~\ref{Tc}. Note that $U=6$ eV and $J=0.6$ eV are used in the following section of this paper. 

As shown in Figure~\ref{LaH10dos}a, the La compound is represented by a two-band system in the absence of a long-range magnetic order in DFT calculation. DFT formalism uses a single Slater determinant model and consequently fails to capture any paramagnetic effects, with an associated magnetic multiplet (fluctuating magnetic moment). These effects usually cause a splitting of spectral features into satellites, as seen in Figure \ref{LaH10dos}b,c, with a resulting enforcement of the $f$ character at the Fermi level. We established that since sharp La features exist at the Fermi level, a greater degree of theoretical modeling is necessary to correctly reflect the superconducting characteristics. In the conducted analysis, for example, the one-shot (DFT+DMFT) and the full-charge self-consistent method (DFT+DMFT+CSC) generate a minor shift of the acute La feature at the Fermi level, which reduces the $f$ character weight at the Fermi level.

According to our results, $f$ orbitals seem to be particularly significant for the superconducting characteristics of rare-earth hydrates. However, the impact of the quantity of the atoms for hydrogen in hydrate with the same $f$-element is unknown. Hence, in addition to LaH$_{10}$, we also studied stable LaH$_{18}$ structures at $400$ GPa with the aim to distinguish any notable tendencies.

A comparison between Figure \ref{LaH18dos}a--c shows that the one-shot (DFT+DMFT) and the full-charge self-consistent method (DFT+DMFT+CSC) barely  change the $4f$La spectral weight at the Fermi level. This leads to similar results for the superconducting temperature of LaH$_{18}$ within DFT, one-shot DFT+DMFT, and DFT+DMFT+CSC approaches. By looking at Figures \ref{LaH10dos} and \ref{LaH18dos}, it can be seen that the $f$ orbital DOS modifications within DMFT with respect to DFT are different for different hydrogen content. For LaH$_{10}$, DFT significantly overestimates  the $4f$-La weight at the Fermi level, while for LaH$_{18}$, the $4f$-La weight is low both in DFT and DMFT, leading to a lower T$_c$.

\section{Methods}
\begin{figure}
\includegraphics[width=1\columnwidth]{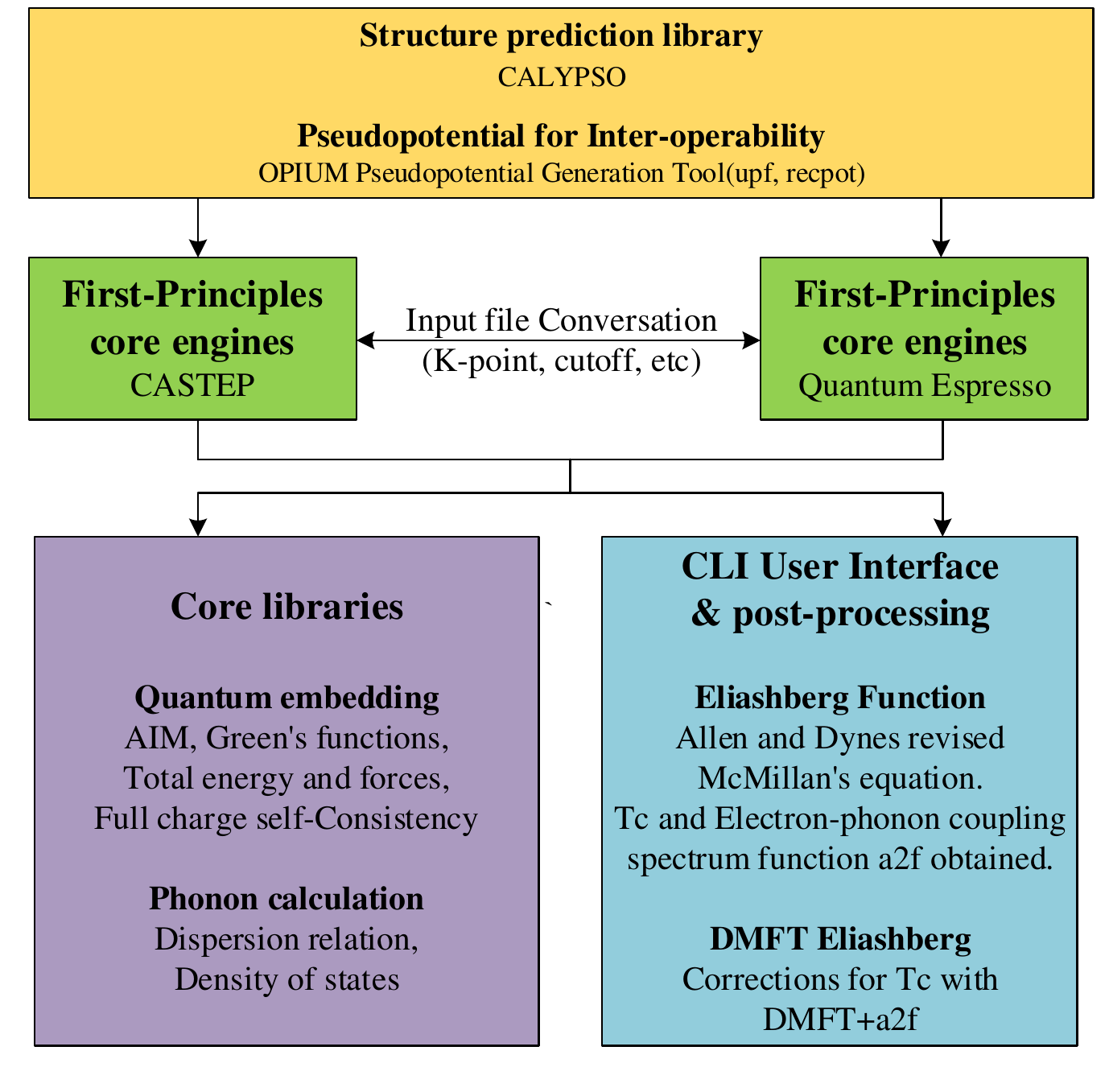}
\caption{\textbf{Schematic of the workflow for DFT+DMFT interfaced with Allen-Dynes}. Overview of the main modules of the theoretical platform and its interrelations. Firstly, structures are predicted by crystal structure analysis by particle swarm optimization (CALYPSO), and the pseudopotentials are generated with OPIUM. The underlying core engines are CASTEP and QE DFT software. We have used format conversion of input files, pseudo-potentials and $\mathbf{k}$-point grids to insure interoperability between QE and CASTEP. Core libraries are employed to provide the many-body corrections, via quantum embedding, which, in turn, provides corrected forces and energies. In post-processing, we used the \textit{DMFT+a2F} approach to obtain values for the Eliashberg function and superconducting temperature $T_c$. Finally, data is archived for future usage using HDF5.}
\label{workflow}

\end{figure}

Our theoretical approach is sketched in Figure~\ref{workflow}. We provide a diagrammatic outline of the main elements of the proposed theoretical platform, as well as their inter-relations. Our technique creates a modular structure for the screening of high-pressure materials.

\textbf{Structural predictions}.
Crystal structure investigation by particle swarm optimization (CALYPSO) \cite{wang2012calypso,wang2010crystal}, based on PSO algorithm~\cite{kennedy1948ieee, eberhat1995new}, provides stoichiometric compositions via Gibbs enthalpies for the equation of state and convex hull. The structure searches were carried out at $400$ GPa with primitive cells of LaH$_{10}$ and LaH$_{18}$ for more than $600$~structures.

\textbf{Ab initio calculations}. Structural optimization and computations of enthalpy were performed using VASP code~\cite{kresse1996efficient}. Electronic structures were calculated by QUANTUM
ESPRESSO (QE)~\cite{giannozzi2009quantum} code.

\textbf{EPC calculations}. We performed the electron-phonon coupling (EPC) calculation using QE with a kinetic energy cutoff of 90 Ry. In order to perform reliable calculation of the electron-phonon coupling in metallic systems, we have employed k-meshes of $2 \times0.045 Å^{-1}$ for the electronic Brillouin zone integration and q-meshes of $2 \times0.09 Å^{-1}$ for LaH$_{10}$ and LaH$_{18}$ compounds.

\textbf{Methods}. We combined the use of QE and CASTEP~\cite{clark2005first,plekhanov2018many} by using input file format conversion, pseudopotentials, and k-point grids.
Core libraries for DMFT quantum embedding were used for the many-body corrections, which provided the total free energies, electronic densities, and Kohn--Sham levels occupancies~\cite{weber2022computational,lee2019mott}.

Norm-conserving pseudopotentials by OPIUM were adopted for describing the core electrons and their effects on the valence orbitals \cite{rappe1991erratum}. Reference electronic configurations of $5s^2 5p^6 5d^1 6s^2$ for the La atoms and $1s^1$ for the H atoms were used. 

\textbf{Post-processing}.
For estimating the superconducting transition temperature, $T_c$ we used the Allen--Dynes modified McMillan equation~\cite{dynes1972mcmillan}:
 
\begin{equation}
T_c = \frac{\omega_{\log}}{1.2} \exp \left [ 
         \frac{-1.04(1+\lambda)}{\lambda - \mu^\star(1+0.62\lambda)}\right ]
\end{equation}
where $\mu^{*}$ is the Coulomb pseudopotential. 
Formulas for the electron-phonon coupling strength $\lambda$ and $\omega_{\log}$ were taken from Ref.~\cite{weber2022computational}. In the post-processing stage, we have followed the procedure outlined by \cite{weber2022computational}; the reader interested in the computation details could refer to this paper for more exhaustive explanations of the steps, equations, and definitions of the relevant terms. Below, we quote the main equations that we used in our calculations of $T_c$.

The Allen--Dynes analysis models the Eliashberg function $\alpha^2F(\omega)$ as a sum at the Fermi level over all scattering processes, mediated by phonon momentum transfer \cite{allen1975transition}:

\begin{equation}
\alpha^2F(\omega) = N(\epsilon_F) \frac{\sum_{\mathbf{k_1},\mathbf{k_2}}{
                    \left|
                    M_{\mathbf{k_1},\mathbf{k_2}}
                    \right|^2
                    \delta(\epsilon_{\mathbf{k_2}})}\delta(\epsilon_{\mathbf{k_1}})
                    \delta(\omega-\omega_{\mathbf{q}\nu})}
                    {\sum_{\mathbf{k_1},\mathbf{k_2}}{
                     \delta(\epsilon_{\mathbf{k_1}}) \delta(\epsilon_{\mathbf{k_2}})}}.
\end{equation}
where $N(\epsilon_F)$ is the DOS at Fermi level, $\omega_{\mathbf{q}\nu}$ is the phonon spectrum of a branch $\nu$ at momentum $\mathbf{q}=\mathbf{k_2}-\mathbf{k_1}$, $\epsilon_{\mathbf{k_1}}$ and $\epsilon_{\mathbf{k_2}}$ are electronic band energies, referred to the Fermi level, while $M_{\mathbf{k_1},\mathbf{k_2}}$ are the electron-phonon coupling matrix elements. When many-body corrections are taken into account, there is a shift of the spectral character at the Fermi level. Furthermore, the electronic correlations lead to a mass enhancement and induce a finite lifetime, due to incoherence. Following onto the DMFT scissors,
the DFT bands are supplied with the renormalized DMFT spectral density:
\begin{equation}
\alpha^2F(\omega) = \mathcal{A}_{tot} \frac{\sum_{\mathbf{k_1},\mathbf{k_2}}{
                    \left|
                    M_{\mathbf{k_1},\mathbf{k_2}}
                    \right|^2
                    \delta(\omega-\omega_{\mathbf{q}\nu}) \mathcal{A}(\mathbf{k_1}) \mathcal{A}(\mathbf{k_2})
                    }}
                    {\sum_{\mathbf{k_1},\mathbf{k_2}}{
                     \mathcal{A}(\mathbf{k_1}) \mathcal{A}(\mathbf{k_2})}},
\end{equation}
where $\mathcal{A}_{tot}$ is the total and $\mathcal{A}(\mathbf{k})$ is the $\mathbf{k}$-momentum resolved spectral weight at the Fermi level. This method is indicated as \textit{DMFT+a2F} in the model diagram. We would like to point out that, unlike the Allen--Dynes method using DFT eigenvalues, which suffers from the double-broadening issue, within our approach, the broadening is given by the finite lifetime from the many-body Green's function and, therefore, our approach is free from the double broadening issue.

The DFT Kohn--Sham Eigenstates are employed in the derivation of the DMFT Green's function in the DFT+DMFT quantum embedding approach \cite{allen1975transition}. The Anderson Impurity Model (AIM) is defined using atomic projectors and solved sequentially inside the Hubbard-I approximation. Furthermore, TRIQS open-source platform~\cite{parcollet2015triqs} provides a wide range of quantum solvers through an interface to DMFT inside CASTEP. The Kohn--Sham potential is computed using the DMFT electronic density, which is generated from the DMFT occupancies, in the complete charge self-consistent technique (DFT+DMFT+CSC). Total energies and forces are determined using Green's function and self-energy after DMFT convergence.
\section{Conclusions}

We used many-body adjustments to predict the superconducting temperature in lanthanide hydrides and explored the significance of this model under different compositions with the same elements. To reestablish a consistent theoretical framework, the DMFT charge self-consistency method was utilized, including many-body corrections to the local charge density in first-principles calculations. Our methodology is free, flexible, and simple to implement, providing interaction of first- principles software with modular foundation. The software includes  QE, CASTEP, and CASTEP+DMFT. 

We found that an increase in the spectral weight of the $f$ states at the Fermi level  led to an increase in the estimated superconducting temperature. In this research, within the DFT+DMFT+CSC method, T$_c$ was changed by 18\% for LaH$_{10}$ and by 3\% for LaH$_{18}$ at $400$~GPa pressure.
Furthermore, we discovered that the adjustment of the superconducting temperature of lanthanide materials within DFT+DMFT+CSC is more significant in symmetric structures and when the fraction of hydrogen elements is not excessively large.


\section*{Data Availability}
The codes are available at url \textbf{dmft.ai} under the GPL 3.0 license. 
\quad

\section*{Author Contribution}
Y.W. carried out the calculations. All authors wrote the paper. C.W. designed the research. All authors have read and agreed to the published version of
the manuscript

\section*{Funding}
Y.W. is supported by the China Scholarship Council. C.W., N.B., and E.P. are supported by Grant EP/R02992X/1 from the U.K. Engineering and Physical Sciences Research Council (EPSRC)

\section*{Acknowledgment}
Y.W. is thankful to Dr. Ying Sun(Jilin University) for valuable discussions. This work was performed using resources provided by the ARCHER U.K. National Supercomputing Service and the Cambridge Service for Data Driven Discovery (CSD3) operated by the University of Cambridge Research Computing Service (www.csd3.cam.ac.uk), provided
by Dell EMC and Intel using Tier-2 funding from the Engineering and Physical Sciences Research
Council (Capital Grant EP/P020259/1), and DiRAC funding from the Science and Technology Facilities Council (www.dirac.ac.uk)

\end{document}